\documentclass[twocolumn,preprintnumbers,amsmath,amssymb]{revtex4}

\usepackage[usenames]{color}

\newcommand{\bea}{\begin{eqnarray}}
\newcommand{\eea}{\end{eqnarray}}

\begin{document}

\title{Evolution of baryon density perturbation in a relativistic MOND model  based on Lorentz-violating vector field}
\author{Jai-chan Hwang${}^{1}$ and Hyerim Noh${}^{2}$}
\address{${}^{1}$Particle Theory  and Cosmology Group,
         Center for Theoretical Physics of the Universe,
         Institute for Basic Science (IBS), Daejeon, 34051, Republic of Korea
         \\
         ${}^{2}$Theoretical Astrophysics Group, Korea Astronomy and Space Science Institute, Daejeon, Republic of Korea
         }


\begin{abstract}

A candidate for relativistic MOND with successful cosmology was proposed recently by using a Lorentz-violating vector field in Einstein's gravity. We show that the dynamic nature of the vector field makes it challenging to realize the MOND. Only in the stationary limit, thus excluding cosmological situations, one can achieve the MOND limit. We study the evolution of density perturbations in the baryon-vector field system using both post-Newtonian approximation and relativistic perturbation theory. Our results show that, considering the dynamic nature of the vector field, the faster growth of structures expected in the low-acceleration MOND regime is not recovered in the relativistic MOND.

\end{abstract}

\maketitle
\tableofcontents

%
%
%
\section{Introduction}
                                        \label{sec:Introduction}

The MOND theory \cite{MOND-1983a}, with stronger gravity in the low-acceleration regime, is gaining ground in cosmology and astrophysics. Recent Gaia observation of wide binary systems can probe the region covering the critical acceleration scale of MOND, $a_M \simeq 1.2 \times 10^{-8} cm/sec^2$. Although there are other opinions, studies show supporting evidence for MOND \cite{Chae-2023, Chae-2024, Chae-2025}; this issue, which may be resolved in near future, reminds us again that, independent of our belief, Newton's gravity, in fact, has never been tested in this low-acceration range including the galactic scale. Recent discovery of high-redshift massive galaxies  \cite{Carniani-2024} is also a favorable observation to MOND as structures can grow faster in the low-acceleration regime with stronger gravity than in the Newtonian gravity \cite{Sanders-1998, McGaugh-2024}. Despite successes in diverse phenomenology, one limiting weak point of MOND in academic practice is the lack of a relativistic theory which allows the cosmological study.

Blanchet, Marsat and Skordis \cite{Blanchet-Marsat-2011, Blanchet-Skordis-2024} recently suggested a simple relativistic extension of a MOND proposal, by Bekenstein and Milgrom \cite{Bekenstein-Milgrom-1984}, with successful cosmology. We call it the BMS theory of relativistic MOND. In this theory, we presented first-order post-Newtonian (1PN) approximation and fully-nonlinear and exact perturbation theory in cosmology \cite{RMOND}. We confirmed that to the linear order perturbation the cosmology is indeed successful as one of the field ($Q$, later) introduced in \cite{Blanchet-Skordis-2024} behaves like the cold dark matter (CDM) in the super-Jeans scale which can be arbitrarily reduced by a parameter choice. However, in the MOND side, BMS theory recovered the Bekenstein and Milgrom's MOND by assuming a stationarity condition \cite{Flanagan-2023, Blanchet-Skordis-2024} which might to be too narrow for cosmological study. 

Here, we study the 0PN formulation and linear perturbation theory of the BMS theory in more details without assuming the stationarity. One important issue in our focus is related to the recent JWST observation of the early emergence of massive galaxies in high redshifts \cite{Carniani-2024}. This observation is often regarded as a challenge to the CDM cosmology. On the other hand, the early formation of galaxies was a predicted feature in the MOND model \cite{Sanders-1998, McGaugh-2024}. In the low-acceleration MOND regime, with  gravity stronger than Newtonian one, not only galatic rotations are sustained without the aid of dark matter \cite{MOND-1983a}, the structures can grow faster \cite{Nusser-2002}, and galaxies can form earlier \cite{Sanders-1998}. In \cite{RMOND} we showed the case in the BMS model by ignoring a certain term ($\sigma$, later) which is in fact possible only in the stationary system. In the following we will show the behavior of baryon density perturbation in the BMS theory without assuming stationarity, in both 0PN and linear perturbation studies.

The BMS's relativistic MOND is based on the following Lorentz-violating combinations of a four-vector $U_a$
\bea
   {\cal L} = \sqrt{-g} \Big\{ {c^4 \over 16 \pi G}
       \big[ R - 2 {\cal J} (A) + 2 {\cal K} (Q) \big]
       + L_{\rm m} \Big\},
   \label{L}
\eea
where
\bea
   & & \hskip -.8cm
       U_a \equiv - {c \over Q} \nabla_a \tau, \quad
       Q \equiv c \sqrt{- (\nabla^c \tau) \nabla_c \tau},
   \nonumber \\
   & & \hskip -.8cm
       A_a \equiv c^2 U_{a;b} U^b = - c^2 q^b_a \nabla_b \ln{Q}, \quad
       A \equiv {1 \over c^4} A^c A_c.
   \label{Q}
\eea
$L_m$ is the matter part including baryon and photons; $U_a$ is a normalized timelike four-vector with $U^c U_c \equiv - 1$; $q_{ab} \equiv g_{ab} + U_a U_b$ is a spatial projection tensor orthogonal to $U_a$; $A_a$ is an acceleration of the four-vector. Our MOND refers to the specific one by Bekenstein and Milgrom \cite{Bekenstein-Milgrom-1984}.

As the gravity part is not modified, if successful, the BMS's MOND theory is accommodated {\it within} Einstein's gravity using a special form of the vector field. ${\cal J}$ can be designed to achieve MOND in the low acceleration regime \cite{Blanchet-Marsat-2011}, and ${\cal K}$ can be designed to handle successful cosmology without dark matter in the large-scale structure, cosmic background radiation and the background cosmology \cite{Blanchet-Skordis-2024, RMOND}.

Previously, we elaborated the BMS theory by studying the PN approximation and the relativistic perturbation theory \cite{RMOND}. We considered $\tau = \tau(t)$ for the background Friedmann cosmology and $\tau = \tau(t) + {1 \over c^2} \sigma ({\bf x}, t)$ for relativistic perturbation theory. However, for the PN analysis we assumed $\tau = t + {1 \over c^2} \sigma ({\bf x}, t) + \dots$, thus ignoring the $\tau$-field contribution to the background cosmology \cite{RMOND}; the PN study in \cite{Blanchet-Marsat-2011, Blanchet-Skordis-2024} considered Minkowski background. This has limited our PN study in the cosmological context by ignoring the $\tau$-field contribution to the background evolution; we can recover the background contribution, though, by replacing $\sigma ({\bf x}, t) \rightarrow \bar \sigma (t) + \sigma ({\bf x}, t)$, and similarly to 1PN order, in \cite{RMOND}.

Here, we extend the PN approximation in cosmology by explicitly taking an {\it ansatz}
\bea
   \tau ({\bf x}, t) = t
       + {1 \over c^2} \bar \sigma (t)
       + {1 \over c^2} \sigma ({\bf x}, t)
       + \dots,
\eea
where $c^{-2}$ order corresponds to the 0PN order. As the MOND modifies gravity in the Newtonian (0PN) level \cite{MOND-1983a, Bekenstein-Milgrom-1984}, in our study we need only 0PN approximation. An overbar indicates background order with $\bar \tau = t + {1 \over c^2} \bar \sigma$ and this can accommodate the background contribution necessary in cosmology. Thus, in the PN approximation we will consider the PN expansion of even the background order $\bar \tau$-field, whereas for the background equations we consider $\bar \tau$-field in exact form. Thus, in the PN approximation, ${1 \over c^2} \bar \sigma$ is one PN order higher than $t$. This allows a consistent study of the PN approximation in cosmology. 

In \cite{Blanchet-Marsat-2011, Blanchet-Skordis-2024, Flanagan-2023}, a condition $\sigma = 0$, which follows by demanding a stationary system, is used to achieve a MOND proposal of \cite{Bekenstein-Milgrom-1984} in the zeroth-order PN (0PN) approximation, see below Eq.\ (\ref{Poisson-sigma}). In the perturbation theory we can take $\sigma \equiv 0$ as a temporal gauge (slicing, hypersurface) condition corresponding to comoving $\tau$-field gauge \cite{RMOND}. However, in the PN analysis $\sigma = 0$ cannot be used to fix the gauge; to 0PN (Newtonian) order the gauge transformation obviously has no room in the analysis. Thus, in the PN analysis, $\sigma = 0$ must be a physical condition (stationarity) imposed on the $\tau$-field.

Now, with the consistent cosmological PN approximation we can properly examine the implication of $\sigma = 0$ condition. We will show that the condition $\sigma = 0$ in the conventional PN approximation leads to an inconsistency in the cosmological context (see Sec.\ \ref{sec:inconsistency}) as it demands the stationary condition as noticed in Minkowski background \cite{Flanagan-2023, Blanchet-Skordis-2024}. We will show that non-vanishing $\sigma$ leads to dynamic equations for the $\tau$-field [see Eqs.\ (\ref{E-conserv-tau}) and (\ref{M-conserv-tau})] and as a result the MOND is not achieved. An effort to remove $\sigma$-part in the $\tau$-field using a coordinate transformation leads to a new 0PN-like approximation which leads to non-Newtonian dynamics for the baryon fluid and the dynamic $\sigma$ still appears in the Poisson's equation (Sec.\ \ref{sec:trouble}).

We will study the evolution of density perturbation of a baryon and $\tau$-field system in the 0PN approximation and in relativistic perturbation theory. Considering the dynamic nature of $\tau$-field we will show that, even in the supposedly MOND regime, the baryon density perturbation evolves differently from the faster growth expected in the MOND proposal \cite{Sanders-1998}. Thus, in the cosmological context the MOND is not realized in the currently considered models in the BMS theory.

Section \ref{sec:BG} presents background equation where only ${\cal K}$ part contributes. Section \ref{sec:PN} presents 0PN study where only ${\cal J}$ part contributes. Section \ref{sec:linear} presents relativistic linear perturbations in the $\sigma = 0$ gauge where both ${\cal J}$ and ${\cal K}$ contribute. Section \ref{sec:discussion} is discussion.

%
%
%
\section{Background}
                                        \label{sec:BG}

We consider a two-component system consisting of the baryon and the $\tau$-field in a flat Friedmann cosmology. Einstein's equation gives
\bea
   & & H^2 = {8 \pi G \over 3 c^2} ( \mu_b
       + \mu_\tau ) + {\Lambda c^2 \over 3},
   \label{BG-1} \\
   & & {\ddot a \over a} = - {4 \pi G \over 3 c^2}
       ( \mu_b + \mu_\tau + 3 p_\tau )
       + {\Lambda c^2 \over 3},
   \label{BG-3}
\eea
where $H \equiv \dot a/a$ and $\mu_i \equiv \varrho_i c^2$ is the energy density with $i = b$ and $\tau$; each component follows $\dot \mu_i + 3 H ( \mu_i + p_i) = 0$; $p_b = 0$ for baryon, thus $\mu_b \propto a^{-3}$.

For the $\tau$-field, we have $Q = \dot \tau = 1 + \dot \sigma/c^2$. From Eq.\ (13) of \cite{RMOND}, we have
\bea
   {8 \pi G \over c^4} \mu_\tau = - {\cal K}
       + Q {\cal K}_{,Q}, \quad
       {8 \pi G \over c^4} p_\tau = {\cal K},
   \label{mu-p-BG}
\eea
thus
\bea
   w_\tau \equiv {p_\tau \over \mu_\tau}
       = { {\cal K} \over - {\cal K} + Q {\cal K}_{,Q} }, \quad
       c_\tau^2 \equiv {\dot p_\tau \over \dot \mu_\tau}
       = { {\cal K}_{,Q} \over Q {\cal K}_{,QQ} }.
   \label{w-cs}
\eea
The equation of motion in Eq.\ (16) of \cite{RMOND} gives ${\cal K}_{,Q} = {I_0 / a^3}$. As the acceleration is a perturbed order, ${\cal J}$ does not affect the background evolution.

We consider a model for ${\cal K}$ \cite{Blanchet-Skordis-2024}
\bea
   {\cal K} = {2 \nu^2 \over n + 1} {\cal K}_{n+1}
       ( Q - 1 )^{n+1},
   \label{model-I}
\eea
with $n = 1, 2, \dots$ Equations (\ref{mu-p-BG})-(\ref{model-I}) give
\bea
   & & Q = 1 +
       \Big( {I_0 \over 2 \nu^2 {\cal K}_{n+1} a^3} \Big)^{1/n}, \quad
       \mu_\tau = {c^4 I_0 \over 8 \pi G a^3}
       {n Q + 1 \over n + 1},
   \nonumber \\
   & & w_\tau = {Q - 1 \over n Q + 1}, \quad
       c_\tau^2 = {Q -1 \over n Q},
   \label{BG-sols}
\eea
where we set $a_0 \equiv 1$. Defining $I_0 \equiv 8 \pi G \mu_{\tau0}/c^4 = 3 \Omega_{\tau 0} H_0^2/c^2$, we have
\bea
   \mu_\tau = \mu_{\tau 0}{1 \over a^3}
       {n Q + 1 \over n + 1}, \quad
       Q = 1 + \Big[ {3 \over 2 a^3}
       \Big( {L_\nu \over L_H} \Big)^2
       \Omega_{\tau 0} \Big]^{1/n},
\eea
where $L_\nu \equiv 1/(\nu \sqrt{{\cal K}_{n+1}})$ and $L_H \equiv c/H_0$; $\mu_{\tau 0}$ and $\Omega_{\tau 0}$ differ from $\mu_\tau$ and $\Omega_\tau$ at present epoch, but the difference is negligible.

For a dust-like behavior of $\tau$-field in the background, authors of \cite{Blanchet-Skordis-2024} demand $w_{\tau*} \leq 0.0164$ at $a_* \sim 10^{-4.5}$. This leads to
\bea
   L_\nu \leq \Big[
       {2 c^2 \over 3 H_0^2 \Omega_{\tau0}}
       \Big( {(n+1) w_{\tau*} \over 1 - n w_{\tau*}} \Big)^n
       a_*^3 \Big]^{1/2} \equiv L_*,
   \label{constraint-BG}
\eea
where with $\Omega_{\tau0} = 0.26$ and $H_0 = 70(km/sec)/Mpc$, thus $L_H = 4.3 Gpc$, the upper limit gives $L_* = 220, 62, 22pc$ for $n = 1, 2, 3$, respectively. The $\tau$-field behaves arbitrarily close to zero-pressure fluid by reducing $L_\nu$.

%
%
%
\section{0PN approximation}
                                        \label{sec:PN}

To 0PN order, our metric convention is \cite{Chandrasekhar-1965, Hwang-Noh-Puetzfeld-2008}
\bea
   g_{00} = - \Big( 1 + {2 \over c^2} \Phi \Big), \quad
       g_{0i} = 0, \quad
       g_{ij} = a^2 \delta_{ij},
   \label{metric-PN}
\eea
where $x^0 = ct$; for the metric to 1PN order in our context of the relativistic MOND, see Eq.\ (21) in \cite{RMOND}. We have
\bea
   Q = 1 + {1 \over c^2} \dot {\bar \sigma}
       - {1 \over c^2} \Xi, \quad
       \Xi \equiv \Phi - \dot \sigma
       + {1 \over 2 a^2} \sigma^{,i} \sigma_{,i},
   \label{Q-0PN}
\eea
and $U_a$ and $A_a$ are the same as in Eq.\ (26) of \cite{RMOND} with $A_i = \Xi_{,i}$; notice that $A_i$ involves the perturbed gravitational potential $\Phi$ which can be used to modify the Poisson's equation, but it also accompanies the $\tau$-field contributions.

For ${\cal K}$ in Eq.\ (\ref{model-I}), only $n = 1$ contributes to 0PN order and is excluded by conflicting constraints from the background in Eq.\ (\ref{constraint-BG}) and from a plausible Poisson's equation to 0PN order \cite{Blanchet-Skordis-2024}. Excluding $n = 1$, for this model, ${\cal K}$-part of the $\tau$-field with $n \ge 2$ does {\it not} contribute to equations in 0PN order.

The energy conservation equation for $\tau$-field and Einstein's equation, respectively, give
\bea
   & & {1 \over a^3} ( a^3 \varrho_\tau )^{\displaystyle{\cdot}}
       - {1 \over a^2} ( \varrho_\tau \sigma^{,i} )_{,i} = 0,
   \label{E-conserv-0PN-tau} \\
   & & {\Delta \over a^2} \Phi
       = 4 \pi G ( \delta \varrho_b + \delta \varrho_\tau ),
   \label{Poisson-0PN}
\eea
where
\bea
   & & \varrho_\tau = \bar \varrho_\tau
       - {1 \over 4 \pi G a^2} ( {\cal J}_{,A} \Xi^{,i} )_{,i}.
   \label{rho-tau-0PN}
\eea
These are Eqs.\ (32), (34) and (38) in \cite{RMOND}; the equation of motion for $\tau$-field also gives Eq.\ (\ref{E-conserv-0PN-tau}) and the momentum conservation equation for $\tau$-field is identically satisfied, see Eq. (35) in \cite{RMOND}. We have $\varrho_\tau = \bar \varrho_\tau + \delta \varrho_\tau$ with $\bar \varrho_\tau = \bar \varrho_{\cal K}$ and $\delta \varrho_\tau = \delta \varrho_{\cal J}$.

Using Eqs.\ (\ref{Q-0PN}) and (\ref{rho-tau-0PN}), Eq.\ (\ref{Poisson-0PN}) gives a modified Poisson's equation
\bea
   & & {1 \over a^2} \nabla \cdot \big[ ( 1 + {\cal J}_{,A} )
       \nabla \Phi \big]
       = 4 \pi G \delta \varrho_b
   \nonumber \\
   & & \qquad
       + {1 \over a^2} \nabla \cdot \Big\{ {\cal J}_{,A}
       \nabla \Big[ \dot \sigma
       - {1 \over 2 a^2} (\nabla \sigma) \cdot \nabla \sigma
       \Big] \Big\}.
   \label{Poisson-sigma}
\eea
This is the main equation determining whether the BMS theory can achieve MOND in the non-relativistic limit; terms with ${\cal J}_{,A}$ are from the acceleration with $A_i = \Xi_{,i}$. In the absence of $\sigma$ term, we achieve the Bekenstein-Milgrom's MOND \cite{Blanchet-Marsat-2011, Sanders-2011, Blanchet-Skordis-2024, RMOND} with

(I) ${\cal J}_{,A} = 0$ in the Newtonian regime,

(II) $1 + {\cal J}_{,A} = x$ with $x \equiv {1 \over a} |\nabla \Phi|/a_M < 1$ in the MOND regime of low acceleration \cite{Bekenstein-Milgrom-1984} where $a_M \simeq 1.2 \times 10^{-8} cm/sec^2$ is the Milgrom's constant \cite{MOND-1983a}.

Therefore, in order to realize the MOND, (i) $\sigma$ terms should disappear, or (ii) the effect of these terms should be suppressed, or at least (iii) these should not interfere the MOND behavior. Although we will show that the effect of $\sigma$ cannot be removed or suppressed, thus failing to achieve the MOND regime in cosmology, we will keep using these terminologies referring to the two regimes.

Although the momentum conservation equation for $\tau$-field is identically valid, using ${\bf v}_\tau \equiv - {1 \over a} \nabla \sigma$, Eqs.\ (\ref{E-conserv-0PN-tau}) and (\ref{rho-tau-0PN}) can be written as
\bea
   & & \hskip -.8cm
       {1 \over a^3} ( a^3 \varrho_\tau )^{\displaystyle{\cdot}}
       + {1 \over a} \nabla \cdot ( \varrho_\tau {\bf v}_\tau ) = 0,
   \label{E-conserv-tau} \\
   & & \hskip -.8cm
       {1 \over a} \nabla \cdot \Big\{ {\cal J}_{,A}
       \Big[ {1 \over a} ( a {\bf v}_\tau
       )^{\displaystyle{\cdot}}
       + {1 \over a} {\bf v}_\tau \cdot \nabla {\bf v}_\tau
       + {1 \over a} \nabla \Phi \Big] \Big\}
   \nonumber \\
   & & \qquad
       \hskip -.8cm
       = - 4 \pi G \delta \varrho_\tau.
   \label{M-conserv-tau}
\eea
These are the energy and (a sort of) momentum conservation equations of the $\tau$-field, revealing dynamic nature of the $\sigma$ field for ${\cal J}_{,A} \neq 0$.

For the baryon, to 0PN order, we have
\bea
   & & {1 \over a^3} ( a^3 \varrho_b )^{\displaystyle{\cdot}}
       + {1 \over a} \nabla \cdot ( \varrho_b {\bf v}_b ) = 0,
   \label{E-conserv-0PN-baryon} \\
   & & {1 \over a} ( a {\bf v}_b )^{\displaystyle{\cdot}}
       + {1 \over a} {\bf v}_b \cdot \nabla {\bf v}_b
       + {1 \over a} \nabla \Phi = 0.
   \label{M-conserv-0PN-baryon}
\eea
These follow from Eqs.\ (36) and (37) in \cite{RMOND}.

Before we discuss the issue of removing $\sigma$ term, we show the behavior of linear density perturbation of the baryon and $\tau$-field system with proper account of $\sigma$. The result will show that the behavior differs from what we expect in the MOND regime [see Eq.\ (\ref{MOND-NL-solution})], thus removing the two possibilities (ii) and (iii) mentioned above Eq.\ (\ref{E-conserv-tau}).

\subsection{Density perturbations}

Here, we show the behavior of linear density perturbation of the baryon and $\tau$-field system with proper account of $\sigma$. For the density and velocity perturbations, we consider linear order deviations from the Friedmann background, thus linear in $\delta_\tau$, $\sigma$, $\delta_b$ and ${\bf v}_b$. However, as the gravitational potential in the MOND regime is characteristically nonlinear, we keep nonlinear order terms involving $\Phi$ together with ${\cal J}_{,A}$.

Equation (\ref{Poisson-0PN}) gives
\bea
   {\Delta \over a^2} \Phi
       = 4 \pi G ( \varrho_b \delta_b
       + \varrho_\tau \delta_\tau ),
   \label{Poisson-lin-0PN}
\eea
where $\delta_i \equiv \delta \varrho_i/\varrho_i$. Using $\Xi$ in Eq.\ (\ref{Q-0PN}), Eqs.\ (\ref{E-conserv-0PN-tau}) and (\ref{rho-tau-0PN}) give
\bea
   & & \dot \delta_\tau = {\Delta \over a^2} \sigma,
   \label{lin-delta-sigma-0PN} \\
   & & {1 \over a^2} \nabla \cdot ( {\cal J}_{,A}
       \nabla \Phi )
       = - 4 \pi G \varrho_\tau \delta_\tau
       + {\cal J}_{,A} {\Delta \over a^2} \dot \sigma.
   \label{lin-tau-0PN}
\eea
In the MOND regime, as ${\cal J}_{,A}$ involves the perturbed potential, for a proper analysis of density perturbation we may need nonlinear treatment, see Sec.\ \ref{sec:inconsistency}.

For the baryon, Eqs.\ (\ref{E-conserv-0PN-baryon}) and (\ref{M-conserv-0PN-baryon}) give
\bea
   \dot \delta_b + {1 \over a} \nabla \cdot {\bf v}_b = 0, \quad
       {1 \over a} ( a {\bf v}_b )^{\displaystyle{\cdot}}
       + {1 \over a} \nabla \Phi = 0,
   \label{lin-b-0PN}
\eea
thus
\bea
   \ddot \delta_b + 2 H \dot \delta_b
       = {\Delta \over a^2} \Phi.
   \label{delta-b-lin-0PN}
\eea
Equations (\ref{Poisson-lin-0PN})-(\ref{lin-tau-0PN}) and (\ref{delta-b-lin-0PN}) give a closed set of density perturbation equations of the baryon-$\tau$ field system. Below we consider solutions.

(A) In the Newtonian regime, with ${\cal J}_{,A} = 0$, Eq.\ (\ref{lin-tau-0PN}) gives $\delta_\tau = 0$, and Eq.\ (\ref{lin-delta-sigma-0PN}) gives $\sigma = 0$. Thus, we have
\bea
   \ddot \delta_b + 2 H \dot \delta_b
       = 4 \pi G \varrho_b \delta_b, \quad
       \delta_\tau = 0.
\eea
Assuming zero-pressure dominated background with $\Lambda = 0$, thus $a \propto t^{2/3}$ and $\Omega_b + \Omega_\tau = 1$, we have solutions $\delta_b \propto t^n$ with $n = {1 \over 6} ( - 1
\pm \sqrt{1 + 24 \Omega_{b0}} )$.

(B) In the MOND regime, with $1 + {\cal J}_{,A} = x < 1$, Eqs.\ (\ref{Poisson-lin-0PN})-(\ref{lin-tau-0PN}) and (\ref{delta-b-lin-0PN}) give
\bea
   & & \hskip -.8cm
       \ddot \delta_b + 2 H \dot \delta_b
       = 4 \pi G ( \varrho_b \delta_b
       + \varrho_\tau \delta_\tau ),
   \label{delta-b-eq-0PN} \\
   & & \hskip -.8cm
       \ddot \delta_\tau + 2 H \dot \delta_\tau
       = 4 \pi G \varrho_b \delta_b
       - {1 \over a^2} \nabla \cdot ( x \nabla \Phi ).
   \label{delta-tau-eq-0PN}
\eea
For $x \ll 1$, ignoring the nonlinear potential term, we have
\bea
   & & \hskip -.8cm
       {1 \over a^5} \big\{ a^2 \big[ a ( a^2 \dot \delta_b
       )^{\displaystyle{\cdot}}
       \big]^{\displaystyle{\cdot}}
       \big\}^{\displaystyle{\cdot}}
       - 4 \pi G \varrho_b \Big[
       {1 \over a^2} ( a^2 \dot \delta_b )^{\displaystyle{\cdot}}
       + 4 \pi G \varrho_\tau \delta_b \Big]
   \nonumber \\
   & & \qquad
       \hskip -.8cm
       = 0.
\eea
Assuming zero-pressure dominated background, we have solutions $\delta_b \propto t^n$ with
\bea
   n = {1 \over 6} \Bigg[ - 1 \pm
       \sqrt{ 1 + 12 \Omega_{b0}
       \Big( 1 \pm
       \sqrt{ 1 + 4 {\Omega_{\tau 0} \over \Omega_{b 0}} }
       \Big) } \Bigg].
   \label{sol-n}
\eea
For $\Omega_{\tau 0} \neq 0$, $n < {2 \over 3}$ for the growing solution; $n= {2 \over 3}$ for $\Omega_{b0} = 1$ and $n$ decreases as $\Omega_{b0} ( = 1 - \Omega_{\tau 0})$ decreases. Thus, the baryon density perturbation does not grow faster even in the MOND regime where a faster growth is expected, see Sec.\ \ref{sec:inconsistency}.

To the linear order in $\sigma$, Eq.\ (\ref{Poisson-sigma}) gives
\bea
   {1 \over a^2} \nabla \cdot \big[ ( 1 + {\cal J}_{,A} )
       \nabla \Phi \big]
       = 4 \pi G \delta \varrho_b
       - {\Delta \over a^2} \dot \sigma.
   \label{Poisson-sigma-linear}
\eea
We can achieve the MOND for $\dot \sigma = 0$. As our analysis of density perturbation shows, in general, the $\sigma$ term cannot be ignored in the MOND regime; ${\bf v}_\tau \equiv - {1 \over a} \nabla \sigma$ plays the role of perturbed velocity of $\tau$-field, see Eqs.\ (\ref{E-conserv-tau}) and (\ref{M-conserv-tau}). 

Below we will show that, in the MOND (thus without the $\sigma$ terms) we expect faster growth of density perturbation. However, in the BMS's relativistic theory, setting $\sigma = 0$ as a physical condition leads to inconsistency in cosmological context, and removing $\sigma$ using a coordinate transformation leads to troublesome results with no resemblence to Newtonian physics

\subsection{Faster growth expected in MOND, and inconsistency of setting $\sigma = 0$ in cosmology}
                                            \label{sec:inconsistency}

The PN analysis in \cite{Blanchet-Marsat-2011, Blanchet-Skordis-2024, RMOND}, in order to achieve the MOND, sets $\sigma = 0$ as a physical condition on the $\tau$-field; this condition is implied by the stationary condition in \cite{Flanagan-2023, Blanchet-Skordis-2024}. With this condition, Eq.\ (\ref{Poisson-sigma}) gives
\bea
   {1 \over a^2} \nabla \cdot \big[
       ( 1 + {\cal J}_{,A} ) \nabla \Phi \big]
       = 4 \pi G \varrho_b \delta_b,
   \label{Poisson-MOND}
\eea
which is the MOND modification of the Poisson's equation by Bekenstein and Milgrom \cite{Bekenstein-Milgrom-1984}, in the cosmological context. To linear order in density and velocity perturbations, Eqs.\ (\ref{E-conserv-0PN-baryon}) and (\ref{M-conserv-0PN-baryon}) give
\bea
   \ddot \delta_b + 2 H \dot \delta_b
       = {\Delta \over a^2} \Phi.
   \label{delta-b-lin}
\eea

In the MOND regime, we have $1 + {\cal J}_{,A} = x$. Although Eq.\ (\ref{Poisson-MOND}) is nonlinear in the perturbed potential, it can be combined with Eq.\ (\ref{delta-b-lin}) to give solutions \cite{Nusser-2002, RMOND}. In the pressure-dominated case with $\Lambda = 0$, we have a solution
\bea
   {1 \over a} | \nabla \Phi | = {3 \over 10} a_M \Omega_b, \quad
       \delta_b = {{3 \over 10} \Omega_b {1 \over a} \Delta \Phi
       \over 4 \pi G \varrho_b a^3} a^2,
   \label{MOND-NL-solution}
\eea
thus, $\delta_b$ grows faster with $\delta_b \propto a^2$ and $\nabla \Phi \propto a$ compared with $\delta_b \propto a$ and $\nabla \Phi \propto a^0$ in the cold dark matter (CDM) case. This faster growth of structure is what we na\"ively expect as the gravity is stronger in the MOND regime \cite{Nusser-2002, RMOND}.

However, Eq.\ (\ref{Poisson-0PN}) gives
\bea
   \delta_\tau = {{10 \over 3} - \Omega_b
       \over 1 - \Omega_b} \delta_b
       \propto a^2.
\eea
Thus, $\dot \delta_\tau \propto \dot \delta_b \neq 0$, and it {\it contradicts} Eq.\ (\ref{lin-delta-sigma-0PN}) which gives $\dot \delta_\tau = 0$ due to $\sigma = 0$ assumed. Therefore, setting $\sigma = 0$ as a physical condition on the $\tau$-field leads to an inconsistency in cosmology. In \cite{Blanchet-Skordis-2024, Flanagan-2023} the authors considered Minkowski background and to achieve MOND used $\sigma = 0$ which is allowed in stationary systems. Here we show that in a cosmological background the condition leads to an inconsistency. Still, we would like to add that current observational supports of MOND are mainly in stationary situations.

\subsection{Troubles in adapted coordinate}
                                               \label{sec:trouble}

In the PN metric in Eq.\ (\ref{metric-PN}), $\sigma$ is not affected by the gauge transformation \cite{RMOND}. In order to achieve MOND by removing the $\sigma$ terms in Eq.\ (\ref{Poisson-sigma}), authors of \cite{Flanagan-2023, Blanchet-Skordis-2024} considered a coordinate transformation $({\bf x}, t) \rightarrow ({\bf x}, t + \sigma/c^2)$, so that in the new coordinate $\tau = t + \bar \sigma/c^2$ has no perturbation; for simplicity, here we ignore $\bar \sigma$ as in \cite{RMOND}, thus $\tau = t$. This was termed an adapted coordinate or unitary gauge \cite{Flanagan-2023, Blanchet-Skordis-2024}. In this new coordinate, however, the PN expansion is not available in the sense that Newtonian dynamics for the baryon and the Poisson's equation are not recovered in the 0PN limit. This was noticed in \cite{Flanagan-2023, Blanchet-Skordis-2024} and here we show the consequence in more details.

Using the coordinate transformation, we have a new 0PN term appearing in $g_{0i}$ with $g_{0i} = {1 \over c} \sigma_{,i}$. Thus, the 0PN metric in adapted coordinate is
\bea
   & & 
       g_{00} = - \Big( 1 + {2 \over c^2} \Phi \Big), \quad
       g_{0i} = {1 \over c} \sigma_{,i},
   \nonumber \\
   & &
       g_{ij} = a^2 \delta_{ij} \Big( 1 - {2 \over c^2} \Psi \Big).
   \label{metric-0PN-adapted}
\eea
We note that $\Psi$ is 1PN order in the PN order counting \cite{Chandrasekhar-1965}; compare with Eq.\ (\ref{metric-PN}). Thus, in our 0PN analysis, we will ignore it. We introduce it only to mention later that, in this new coordinate, $\Psi$ differs from $\Phi$ in contrast to Newtonian case where $\Psi = \Phi$ to 0PN order \cite{Chandrasekhar-1965}, see below; in the ordinary PN analysis, although $\Psi$ does not affect the Poisson's and conservation equations in the Newtonian limit, it does affect the light propagation, causing $2$ factor difference in the gravitational lensing. 

The inverse metric, connection and curvature, we need, are
\bea
   & & g^{00} = - \Big( 1 - {2 \over c^2} \Phi \Big), \quad
       g^{0i} = {1 \over a^2 c} \sigma^{,i}, \quad
       g^{ij} = {1 \over a^2} \delta^{ij};
   \nonumber \\
   & & \Gamma^0_{00} \sim {\cal O}(c^{-3}), \quad
       \Gamma^0_{0i} = {1 \over c^2} ( \Phi + H \sigma )_{,i},
   \nonumber \\
   & & \Gamma^0_{ij} = {1 \over c} ( a^2 H \delta_{ij}
       - \sigma_{,ij} ), \quad
       \Gamma^i_{00} = {1 \over c^2 a^2}
       ( \Phi + \dot \sigma )^{,i},
   \nonumber \\
   & & \Gamma^i_{0j} = {1 \over c} H \delta^i_j, \quad
       \Gamma^i_{jk} \sim {\cal O} (c^{-2});
   \nonumber \\
   & & R^0_0 = - R_{00} = {1 \over c^2} \Big[ 3 \dot H + 3 H^2
       - {\Delta \over a^2} ( \Phi + \dot \sigma ) \Big].
\eea

For the baryon, we have $T_{ab} = \mu_{b} u_a u_b$ where $u_a$ is the baryon velocity four-vector. In our new metric, we have
\bea
   u_i \equiv {a \over c} v_i, \quad
       u_0 = -1, \quad
       u^i = {1 \over ac} \Big( v^i
       - {1 \over a} \sigma^{,i} \Big), \quad
       u^0 = 1,
\eea
where $v_i$ is the baryon velocity with the index associated with $\delta_{ij}$. Thus,
\bea
   & & \hskip -.5cm
       T^0_0 = - \varrho_b c^2, \quad
       T^0_i = a \varrho_b c v_i, \quad
       T^i_0 = - {c \over a} \varrho_b \Big( v^i
       - {1 \over a} \sigma^{,i} \Big),
   \nonumber \\
   & & \hskip -.5cm
       T^i_j = \varrho_b \Big( v^i
       - {1 \over a} \sigma^{,i} \Big) v_j.
\eea
Conservation equations, $T^b_{a;b} = 0$, for the baryon give
\bea\
   & & \hskip -.8cm
       {1 \over a^3} ( a^3 \varrho_b )^{\displaystyle{\cdot}}
       + {1 \over a} ( \varrho_b v^i )_{,i}
       = {1 \over a^2} ( \varrho_b \sigma^{,i} )_{,i},
   \label{E-conserv-0PN-adapted} \\
   & & \hskip -.8cm
       {1 \over a} ( a v_i )^{\displaystyle{\cdot}}
       + {1 \over a} v^j v_{i,j}
       + {1 \over a} \Phi_{,i}
   \nonumber \\
   & & \qquad
       \hskip -.8cm
       = - {2 \over a} H \sigma_{,i}
       + {1 \over a} \sigma_{,ij} \Big( v^j
       - {1 \over a} \sigma^{,j} \Big)
       + {1 \over a} v_{i,j} \sigma^{,j},
   \label{M-conserv-0PN-adapted}
\eea
which are heavily affected by the presence of $\sigma$ in the metric. Failure in recovering the proper Newtonian conservation equations for an ordinary fluid is a serious drawback as a PN approximation.

Einstein's equation gives
\bea
   R^0_0 = - {4 \pi G \over c^2} ( \varrho_b + \varrho_\tau )
       + \Lambda.
\eea
Subtracting the background equation, we have the Poisson's equation
\bea
   {\Delta \over a^2} (\Phi + \dot \sigma)
       = 4 \pi G ( \delta \varrho_b + \delta \varrho_\tau ).
\eea
Using the revised metric, we have
\bea
   & & \hskip -.5cm
       U_0 = - \Big( 1 + {\Phi \over c^2} \Big), \quad
       U_i = 0, \quad
       U^0 = 1 - {\Phi \over c^2},
   \nonumber \\
   & & \hskip -.5cm
       U^i = - {1 \over a^2 c} \sigma^{,i}; \quad
       A_0 = {1 \over a^2 c} \sigma^{,i} \Phi_{,i}, \quad
       A_i = \Phi_{,i}, \quad
       A^0 = 0,
   \nonumber \\
   & & \hskip -.5cm
       A^{,i} = {1 \over a^2} \Phi^{,i}; \quad
       A = {1 \over c^4 a^2} \Phi^{,i} \Phi_{,i}; \quad
       Q = 1 - { \Phi \over c^2},
\eea
and
\bea
   \delta \varrho_\tau
       = - {1 \over 4 \pi G a^2}
       ( {\cal J}_{,A} \Phi^{,i} )_{,i}.
\eea
Compared with Eq.\ (\ref{rho-tau-0PN}), $\sigma$ terms disappeared in $\delta \varrho_\tau$. However, the Poisson's equation becomes
\bea
   {1 \over a^2} \nabla \cdot \big[ ( 1 + {\cal J}_{,A} )
       \nabla \Phi \big]
       = 4 \pi G \delta \varrho_b
       - {\Delta \over a^2} \dot \sigma,
\eea
which is the same as Eq.\ (\ref{Poisson-sigma-linear}), and still involves $\dot \sigma$ term due to the presence of $\sigma$ in the new 0PN metric. Thus, the adaptive coordinate does not help recovering the MOND. 

Although these are enough to show the  ill-suited nature of the adapted coordinate, further consideration of other part of Einstein's equation reveals that $\Psi$ differs from $\Phi$ to 0PN order which is in contrast to the ordinary PN approximation \cite{Chandrasekhar-1965}.

%
%
%
\section{Relativistic perturbations}
                                        \label{sec:linear}

In order to show the evolution of density perturbation in the BMS theory, we consider the relativistic linear perturbations of the baryon and $\tau$-field system \cite{RMOND}. Conservation equations in Eqs.\ (89) and (90) in \cite{RMOND} give
\bea
   & & \hskip -.8cm
       \dot \delta_b
       = \kappa - 3 H \alpha + {\Delta \over a} v_b,
   \label{eq6-b-linear} \\
   & & \hskip -.8cm
       {1 \over a} ( a v_b )^{\displaystyle{\cdot}}
       = {c^2 \over a} \alpha,
   \label{eq7-b-linear} \\
   & & \hskip -.8cm
       \delta \dot \mu_\tau
       + 3 H ( \delta \mu_\tau + \delta p_\tau )
       = ( \mu_\tau + p_\tau ) \Big( \kappa - 3 H \alpha
       + {\Delta \over a} v_\tau \Big),
   \label{eq6-tau-linear} \\
   & & \hskip -.8cm
       {1 \over a^4} [ a^4 ( \mu_\tau + p_\tau ) v_\tau
       ]^{\displaystyle{\cdot}}
       = {c^2 \over a} [ ( \mu_\tau + p_\tau ) \alpha
       + \delta p_\tau ].
   \label{eq7-tau-linear}
\eea
Einstein's equation to the linear order is presented in Eqs.\ (83)-(87) of \cite{RMOND}. We only need the Raychaudhury equation
\bea
   & & \hskip -.8cm
       \dot \kappa + 2 H \kappa
       + \Big( c^2 {\Delta \over a^2} + 3 \dot H \Big) \alpha
       = {4 \pi G \over c^2}
       ( \delta \mu_b + \delta \mu_\tau + 3 \delta p_\tau ).
   \label{eq4-linear}
\eea
For the $\tau$-field, we have
\bea
   & & \delta \mu_\tau = {c^4 \over 8 \pi G}
       \Big( Q {\cal K}_{,QQ} \delta Q
       - 2 {\cal J}_{,A} {\Delta \over a^2} \Upsilon \Big),
   \nonumber \\
   & &
       \delta p_\tau = {c^4 \over 8 \pi G} {\cal K}_{,Q} \delta Q, \quad
       v_\tau = {1 \over a Q} \sigma,
   \label{fluid-pert}
\eea
with
\bea
   \delta Q = - Q \alpha + {1 \over c^2} \dot \sigma, \quad
       \Upsilon = \alpha
       - {1 \over c^2 Q} \Big( \dot \sigma
       - {\dot Q \over Q} \sigma \Big),
\eea
derived in Eqs.\ (73) and (80) of \cite{RMOND}. Although ${\cal J}_{,A}$ contains perturbed potential in the MOND regime, we {\it regard} it as a coefficient in $\delta \mu_\tau$.

Now, we take $\sigma = 0$ gauge, thus $v_\tau = 0$; thus it is comoving $\tau$-field gauge. We have
\bea
   {\delta p_\tau \over \mu_\tau}
       = c_\tau^2 \Big( 1 - {\cal J}_{,A}
       {c_\tau^2 c^2 k^2 \over 4 \pi G \varrho_\tau a^2}
       \Big)^{-1} \delta_\tau,
   \label{delta-p-mu}
\eea
where we used a model in Eq.\ (\ref{model-I}) with $Q \simeq 1$ and $\Delta = - k^2$. Equations (\ref{eq6-tau-linear}) and (\ref{eq7-tau-linear}) give
\bea
   \kappa = { (a^3 \delta \mu_\tau
       )^{\displaystyle{\cdot}}
       \over a^3 ( \mu_\tau + p_\tau )}, \quad
       \alpha = - {\delta p_\tau \over \mu_\tau + p_\tau}.
   \label{tau-pert-eqs}
\eea

\subsection{Jeans scale}

Considering the $\tau$-field only, assuming near pressureless background for $\tau$-field, Eqs.\ (\ref{eq4-linear}), (\ref{delta-p-mu}) and (\ref{tau-pert-eqs}) give
\bea
   \ddot \delta_\tau
       + 2 H \dot \delta_\tau
       - 4 \pi G \varrho_\tau \delta_\tau
       { {1 \over c_\tau^2} {4 \pi G \varrho a^2 \over c^2 k^2}
       - ( 1 + {\cal J}_{,A} ) \over
       {1 \over c_\tau^2} {4 \pi G \varrho a^2 \over c^2 k^2}
       + ( - {\cal J}_{,A} )} = 0.
\eea
The perturbed pressure term is arranged as a modifying factor in the gravity term. This factor vanishes when the pressure term is comparable to the gravity term, and gives the Jeans scale $\lambda_J \equiv 2 \pi a/k_J$ with \cite{RMOND}
\bea
   \hskip -.05cm
   {a \over k_J}
       = c_\tau c \sqrt{1 + {\cal J}_{,A} \over 4 \pi G \varrho_\tau}
       = \sqrt{ L_\nu^{2/n} \Big( {2 \over 3}
       {L_H^2 a^3 \over \Omega_{\tau 0}} \Big)^{1-1/n}
       {1 + {\cal J}_{,A} \over n} }.
   \label{Jeans}
\eea
For $n = 2$, we have
\bea
   \hskip -.1cm
   {a \over k_J}
       = \sqrt{ {L_H L_\nu a^{3/2} ( 1 + {\cal J}_{,A} )
       \over \sqrt{6 \Omega_{\tau0}}}}
       \le \sqrt{1 + {\cal J}_{,A}} a^{3/4} .46 Mpc.
   \label{constraint-pert}
\eea
In the second step we used the background constraint in Eq.\ (\ref{constraint-BG}). The Jeans scale can become arbitrarily small by reducing $L_\nu$, thus making the $\tau$-field behave as a pressureless dark matter, like the CDM, to the linear order.

\subsection{Density perturbations}

Now, we consider the two-component system in the $\sigma = 0$ gauge. For the baryon we have Eqs.\ (\ref{eq6-b-linear}) and (\ref{eq7-b-linear}). For the $\tau$-field, we have Eq.\ (\ref{tau-pert-eqs}). Equation (\ref{eq4-linear}) gives
\bea
   & & {1 \over a^2} ( a^2 \kappa )^{\displaystyle{\cdot}}
       = 4 \pi G ( \varrho_b \delta_b
       + \varrho_\tau \delta_\tau )
   \nonumber \\
   & & \qquad
       + \Big( c^2 {\Delta \over a^2} - 12 \pi G \varrho_b \Big)
       {\delta p_\tau \over \mu_\tau + p_\tau},
   \label{kappa-pert-eq}
\eea
with $\delta p_\tau$ given in Eq.\ (\ref{delta-p-mu}). Combining these, we have
\bea
   & & {1 \over a^2} ( a^2 \dot \delta_b
       )^{\displaystyle{\cdot}}
       - 4 \pi G \Big( \varrho_b \delta_b
       + \varrho_\tau \delta_\tau
       - 3 \varrho_b {\delta p_\tau
       \over \mu_\tau + p_\tau} \Big)
   \nonumber \\
   & & \qquad
       = {3 \over a^3} \Big( {a^2 H \delta p_\tau \over
       \mu_\tau + p_\tau} \Big)^{\displaystyle{\cdot}},
   \label{delta-b-eq-pert} \\
   & & {1 \over a^2} \Big[ {( a^3 \delta \mu_\tau
       )^{\displaystyle{\cdot}} \over a ( \mu_\tau + p_\tau)}
       \Big]^{\displaystyle{\cdot}}
       - 4 \pi G \Big( \varrho_b \delta_b
       + \varrho_\tau \delta_\tau
       - 3 \varrho_b {\delta p_\tau
       \over \mu_\tau + p_\tau} \Big)
   \nonumber \\
   & & \qquad
       = c^2 {\Delta \over a^2}
       {\delta p_\tau \over \mu_\tau + p_\tau}.
   \label{delta-tau-eq-pert}
\eea
Assuming near pressureless background for $\tau$-field, we have
\bea
   & & \hskip -.8cm
       \ddot \delta_b + 2 H \dot \delta_b
       - 4 \pi G \Big( \varrho_b \delta_b
       + \varrho_\tau \delta_\tau
       - 3 \varrho_b {\delta p_\tau \over \mu_\tau} \Big)
   \nonumber \\
   & & \qquad
       \hskip -.8cm
       = {3 \over a^3} \Big( a^2 H {\delta p_\tau \over
       \mu_\tau} \Big)^{\displaystyle{\cdot}},
   \\
   & & \hskip -.8cm
       \ddot \delta_\tau + 2 H \dot \delta_\tau
       - 4 \pi G \Big( \varrho_b \delta_b
       + \varrho_\tau \delta_\tau
       - 3 \varrho_b {\delta p_\tau \over \mu_\tau} \Big)
       = c^2 {\Delta \over a^2}
       {\delta p_\tau \over \mu_\tau}.
   \label{delta-tau-eq-pert-2}
\eea
Below we consider solutions.

(A) In the Newtonian regime, with ${\cal J}_{,A} = 0$, we have $\delta p_\tau / \mu_\tau = c_\tau^2 \delta_\tau$ with $c_\tau^2 \ll 1$, and
\bea
   & & \hskip -.8cm
       \ddot \delta_b + 2 H \dot \delta_b
       - 4 \pi G ( \varrho_b \delta_b
       + \varrho_\tau \delta_\tau ) = 0,
   \\
   & & \hskip -.8cm
       \ddot \delta_\tau + 2 H \dot \delta_\tau
       - 4 \pi G ( \varrho_b \delta_b
       + \varrho_\tau \delta_\tau )
       = c_\tau^2 c^2 {\Delta \over a^2} \delta_\tau.
\eea
Above the Jeans scale, the $\tau$-field behaves as a CDM.

(B) In the super-Jeans scale with $(k/k_J)^2 \ll 1$, for general ${\cal J}_{,A}$, we have $\delta p_\tau/\mu_\tau \sim c_\tau^2 \delta_\tau$, and
\bea
   & & \hskip -.8cm
       \ddot \delta_b + 2 H \dot \delta_b
       - 4 \pi G ( \varrho_b \delta_b
       + \varrho_\tau \delta_\tau ) = 0,
   \\
   & & \hskip -.8cm
       \ddot \delta_\tau + 2 H \dot \delta_\tau
       - 4 \pi G ( \varrho_b \delta_b
       + \varrho_\tau \delta_\tau ) = 0.
\eea
Thus, the density perturbation of $\tau$-field behaves as a CDM. Assuming zero-pressure dominated background with $\Lambda = 0$, the solutions are $\delta_b \propto \delta_\tau \propto (t^{2/3}, t^0, t^{-1/3}, t^{-1})$; for the growing solution, we have $\delta_b = \delta_\tau \propto t^{2/3}$.

(C) In the sub-Jeans scale with $(k/k_J)^2 \gg 1$, we have $\delta p_\tau/\mu_\tau = - (1 + 1/{\cal J}_{,A}) c_\tau^2 (k_J / k)^2 \delta_\tau \ll \delta_\tau$. In the MOND regime with $1 + {\cal J}_{,A} = x$, we have
\bea
   & & \hskip -.8cm
       \ddot \delta_b + 2 H \dot \delta_b
       - 4 \pi G ( \varrho_b \delta_b
       + \varrho_\tau \delta_\tau ) = 0,
   \\
   & & \hskip -.8cm
       \ddot \delta_\tau + 2 H \dot \delta_\tau
       - 4 \pi G \Big( \varrho_b \delta_b
       - {x \over 1 - x}
       \varrho_\tau \delta_\tau \Big) = 0.
\eea
These are consistent with Eqs.\ (\ref{delta-b-eq-0PN}) and (\ref{delta-tau-eq-0PN}) with solutions for $x \ll1$ in Eq.\ (\ref{sol-n}). Thus, properly considering the dynamic nature of $\tau$-field, no faster growth of baryon perturbation can be achieved in the MOND regime. This implies that due to the dynamic nature of the $\tau$-field the BMS theory fails to reproduce the faster growth expected in the MOND.

%
%
%
\section{Discussion}
                                        \label{sec:discussion}

We examined a relativistic MOND model based on a Lorentz-violating vector field \cite{Blanchet-Marsat-2011, Blanchet-Skordis-2024}. Compared with our previous work on the subject in \cite{RMOND} we made a consistent PN approximation properly accommodating background cosmology. This allows a consistent cosmological study using the PN approximation.

Using the consistent formulation of the cosmological PN approximation we show that the dynamic nature of the $\tau$-field causes difficulty in reproducing Bekenstein-Milgrom's MOND theory in the BMS model. Suppressing the dynamic part by simply ignoring it leads to inconsistency in cosmology. Aligning the $\tau$-field to the background cosmic time $t$ by a coordinate transformation causes reformulation of the 0PN metric with the perturbed $\tau$ part appearing in the metric. This causes non-Newtonian behavior of conservation equations for the baryon, and the perturbed $\tau$ part appears again in the Poisson's equation, thus still difficult to reproduce the MOND.

We studied the evolution of density perturbations of the baryon and $\tau$-field system, in both the 0PN approximation and the relativistic perturbation theory, the latter in $\sigma = 0$ gauge. The PN approximation and relativistic perturbation theory are complementary but entirely different approach to probe the relativistic aspects of Einstein's gravity; the former is weakly relativistic but fully nonlinear, whereas the latter is fully relativistic but weakly nonlinear.

In our case, there are subtle differences in results between the two approaches. To 0PN order, while ${\cal K}$ part does not contribute to the energy density, pressure and stress, ${\cal J}$ part contributes to all these fluid variables, see Eq.\ (32) of \cite{RMOND}. This can be compared with the perturbation theory where to the linear order we have non-vanishing pressure coming from the ${\cal K}$ part while the stress vanishes. Thus, we have Jeans scale for $\tau$-field in perturbation theory in contrast to the 0PN approximation. Although the MOND issue appears in the Newtonian (0PN) limit, for a proper comparison of the linear perturbation equations in PN approximation with relativistic perturbation theory, we need up to 1PN approximation which also involves relativistic linear order terms. For a comparison in a single-component zero-pressure fluid, see \cite{NH-PN-2012}.

Due to stronger gravity in the low-acceleration MOND regime the baryonic structures can grow faster than in Newton's gravity, see Eq.\ (\ref{MOND-NL-solution}). The early emergence of galaxies was a predicted feature in the MOND model \cite{Sanders-1998, McGaugh-2024}, and is in accord with recent JWST observation of the early presence of massive galaxies in high redshifts \cite{Carniani-2024}; this phenomenon is often regarded as a challenge to the CDM paradigm. In both PN approximation and relativistic perturbation theory, considering $\sigma$ term properly, we showed that the $\tau$-field recovers CDM behavior in the super-Jeans scale. However, in the BMS theory, we found {\it no} faster growth of the baryon density perturbation in the MOND regime. This reveals another aspect that the current version of BMS theory has difficulty in reproducing the MOND paradigm.

\vskip .5cm
%
%
%
\centerline{\bf Acknowledgments}

We wish to thank Professors L. Blanchet and E. Flanagan for insightful comments and suggestions. H.N.\ was supported by the National Research Foundation (NRF) of Korea funded by the Korean Government (No.RS-2024-00333721 and No.2021R1F1A1045515). J.H.\ was supported by IBS under the project code, IBS-R018-D1.

%
%


\end{document}